\documentclass{ws-ijbc}
\usepackage{graphicx}
\usepackage{epstopdf}
\DeclareMathOperator{\sat}{sat}
\begin{document}

\catchline{}{}{}{}{} 

\markboth{N.V.~Stankevich, N.V.~Kuznetsov, G.A.~Leonov, L.~Chua}{Scenario of the birth of hidden attractors in the Chua circuit}

\title{Scenario of the Birth of Hidden Attractors in the Chua Circuit}

\author{Nataliya V. Stankevich}

\address{
Saint-Petersburg State University, Russia; \\
Yuri Gagarin State Technical University of Saratov, Russia; \\
University of Jyv\"{a}skyl\"{a}, Finland;
\\ stankevichnv@mail.ru}

\author{Nikolay V. Kuznetsov}
\address{
Saint-Petersburg State University, Russia; \\
University of Jyv\"{a}skyl\"{a}, Finland;
\\ nkuznetsov239@gmail.com}

\author{Gennady A. Leonov}
\address{
Saint-Petersburg State University, Russia
\\ g.leonov@spbu.ru}

\author{Leon O. Chua}
\address{
University of California, USA
\\ chua@eecs.berkeley.edu}

\maketitle

\begin{history}
\received{(to be inserted by publisher)}
\end{history}

\begin{abstract}
Recently it was shown that in the dynamical model of Chua circuit
both the classical self-excited and hidden chaotic attractors
can be found.
In this paper the dynamics of the Chua circuit is revisited.
The scenario of the chaotic dynamics development
and the birth of self-excited and hidden attractors is studied.
It is shown a  pitchfork bifurcation
in which a pair of symmetric attractors coexists and merges
into one symmetric attractor through an attractor-merging bifurcation
and a splitting of a single attractor into two attractors.
The scenario relating the subcritical Hopf bifurcation near equilibrium
points and the birth of hidden attractors is discussed.
\end{abstract}


\section{Introduction}
The Chua circuit is one of the well-known and well-studied
nonlinear dynamical models
\cite{Chua2-1992,Chua-1992,KuznetsovKSC-1993,BelykhC-1993,NekorkinC-1993,LoziU-1993,ShilnikovTCh-2001,BilottaP-2008}.
To date in the Chua circuit it has been found chaotic attractors of various
shapes (see, e.g. a gallery of Chua attractors in \cite{BilottaP-2008}).
Until recently, all the known Chua attractors were self-excited attractors,
which can be numerically visualized by a trajectory starting from
a point in small neighborhood of an unstable equilibrium.

{\bf Definition.}
{\cite{LeonovKV-2011-PLA,LeonovK-2013-IJBC,LeonovKM-2015-EPJST,Kuznetsov-2016}}
{\it  An attractor is \emph{hidden}
 if its basin of attraction does not intersect
 with a neighborhood of all equilibria (stationary points);
 otherwise, it is called a \emph{self-excited} attractor.
}

For a \emph{self-excited attractor}, its basin of attraction
is connected with an unstable equilibrium
and, therefore, self-excited attractors
can be localized numerically by the
\emph{standard computational procedure}
in which after a transient process a trajectory,
started in a neighborhood of an unstable equilibrium
(e.g., from a point of its unstable manifold),
is attracted to a state of oscillation and then traces it.
Thus, self-excited attractors can be easily visualized
(e.g., the classical Lorenz, R\"{o}ssler,
and Hennon  attractors can be visualized
by a trajectory from a vicinity of unstable zero equilibrium).

For a hidden attractor, its basin of attraction is not connected with
equilibria, and, thus, the search and visualization of hidden attractors
in the phase space may be a challenging task.
Hidden attractors are attractors in systems without equilibria
(see, e.g. rotating electromechanical systems with Sommerfeld effect
described in 1902 \cite{Sommerfeld-1902,KiselevaKL-2016-IFAC})
and in systems with only one stable equilibrium
(see, e.g. counterexamples \cite{LeonovK-2011-DAN,LeonovK-2013-IJBC}
to the Aizerman's (1949) and Kalman's (1957) conjectures
on the monostability of  nonlinear control systems \cite{Aizerman-1949,Kalman-1957}).
One of the first related problems is the second part of Hilbert's 16th problem (1900)
\cite{Hilbert-1901}
on the number and mutual disposition of limit cycles
in two-dimensional polynomial systems, where nested limit cycles
(a special case of multistability and coexistence of attractors)
exhibit hidden periodic oscillations
(see, e.g., \cite{Bautin-1939,KuznetsovKL-2013-DEDS,LeonovK-2013-IJBC}).

The \emph{classification of attractors as being hidden or self-excited}
was introduced by G.~Leonov and N.~Kuznetsov
in connection with the discovery of the first hidden Chua attractor \cite{LeonovK-2009-PhysCon,KuznetsovLV-2010-IFAC,LeonovKV-2011-PLA,BraginVKL-2011,LeonovKV-2012-PhysD,KuznetsovKLV-2013,LeonovK-2013-IJBC,LeonovKKK-2015-IFAC}
and has captured attention of scientists from around the world
(see, e.g. \cite{BurkinK-2014-HA,LiSprott-2014-HA,Chen-2015-IFAC-HA,SahaSRC-2015-HA,FengP-2017-HA,ZhusubaliyevMCM-2015-HA,Danca-2016-HA,KuznetsovKMS-2015-HA,ChenLYBXW-2015-HA,PhamRFF-2014-HA,OjoniyiA-2016-HA,RochaM-2016-HA,BorahR-2017-HA,DancaKC-2016,WeiPKW-2016-HA,PhamVJVK-2016-HA,JafariPGMK-2016-HA,DudkowskiJKKLP-2016,SinghR-2017-HA,ZhangWWM-2017-HA,MessiasR-2017-HA,BrzeskiWKKP-2017,WeiMSAZ-2017-HA,ChaudhuriP-2014-HA,JiangLWZ-2016-HA,VolosPZMV-2017-HA}).

Further study of the hidden Chua attractors and their
observation in physical experiments can be found, e.g. in \cite{LiZY-2014-HA,ChenLYBXW-2015-HA,BaoHCXY-2015-HA,ChenYB-2015-HA,ChenYB-2015-HA-EL,Zelinka-2016-HA,BaoLWX-2016,MenacerLC-2016-HA,ChenKLM-2017-IJBC,RochaRK-2017-HA,HlavackaG-2017-HA}.
The synchronization of Chua circuits with hidden attractors is discussed,
e.g. in \cite{KuznetsovL-2014-IFACWC,KuznetsovLMS-2016-INCAAM,KuznetsovLS-2017,KiselevaKKKLYY-2017}.
Also some recent results on various modifications of Chua circuit can be found in
\cite{RochaM-2015,BaoJWH-2015,Semenov20151553,GribovKK-2016,Kengne-2016,ZhaoLD-2017-HA,ChenXLB-2017,CorintoF-2017}.


In this work the scenario of the chaotic dynamics development
and the birth of self-excited and hidden Chua attractors is studied.
It is shown a pitchfork bifurcation
in which a pair of symmetric attractors coexists and merges
into one symmetric attractor through an attractor-merging bifurcation
and a splitting bifurcation of a single attractor into two attractors.
It is presented the scenario of the birth of hidden attractor connected with a
subcritical Hopf bifurcation near equilibrium points and a saddle-node bifurcation of a limit cycles.
In general, the \emph{conjecture is that for a globally bounded autonomous system
of ordinary differential equations
with unique equilibrium point, which is
asymptotically stable, the subcritical Hopf bifurcation
leads to the birth of a hidden attractor.}\footnote{
The conjecture was formulated in 2012 by L.~Chua
in private communication with N.~Kuznetsov and G.~Leonov.}

\section{Dynamical regimes of the Chua circuit}\label{sec:2}
The Chua circuit, invented in 1983 by Leon Chua \cite{Chua2-1992, Chua-1992}, 
is the simplest electronic circuit exhibiting chaos.
The classical Chua circuit can be described by the following differential equations
\begin{equation}
\label{Chua}
  \begin{array}{l l}
    \dot {x}=\alpha(y-x)-\alpha f(x),\\
    \dot {y}=x-y+z,\\
    \dot {z}=-(\beta y+\gamma z),
  \end{array}
\end{equation}
where $f(x)=m_{1}x+\frac{1}{2} (m_{0}-m_{1})(|x+1|-|x-1|)$ is
a  piecewise linear  voltage-current characteristic.
Here $x$, $y$, $z$ are dynamical variables;
parameters $m_{0}$, $m_{1}$ characterize a  piecewise linear characteristic of
nonlinear element;
parameters $\alpha$, $\beta$, and $\gamma$ characterize a resistor, a capacitors,
and an inductance.
It is well known that model \eqref{Chua} is symmetric with respect to
the origin and  remains unchanged under the transformation
($x, y, z) \to (-x, -y, -z)$.

System \eqref{Chua} can be considered as
a feedback control system in the Lur'e form
\begin{equation} \label{Chua-2}
\begin{aligned}
    & \dot {u}=P{u}+q\phi(r^*{u}), \quad {u}=(x,y,z) \in \mathbb{R}^3, \\
    & P=\left(
               \begin{array}{ccc}
                 -\alpha (m_{1}+1) & \alpha & 0 \\
                 1 & -1 & 1 \\
                 0 & -\beta & -\gamma \\
               \end{array}
             \right), \
       q=\left(
              \begin{array}{c}
              -\alpha \\
               0 \\
               0
              \end{array}
      \right), \
      r=\left(
              \begin{array}{c}
               1 \\
               0 \\
               0
              \end{array}
    \right), \\
    & \phi(x)=(m_{0}-m_{1})\sat(x) = \frac{1}{2} (m_{0}-m_{1})(|x+1|-|x-1|).
\end{aligned}
\end{equation}

\subsection{Local analysis of equilibrium points}
\label{sec:2.1}
Suppose that
\begin{equation}
\label{lim}
  \begin{array}{l l}
(\beta\neq-\gamma) \text{\ and }
\bigg(
\big(m_0<-\frac{\beta}{\gamma+\beta} \text{\ and } m_1>-\frac{\beta}{\gamma+\beta}\big)
\text{\ or }
\big(m_1<\frac{1}{2}(m_0-\frac{\beta}{\gamma+\beta}) \text{\ and } m_1>-\frac{\beta}{\gamma+\beta}\big)
\text{\ or } \\
\big(m_0>-\frac{\beta}{\gamma+\beta} \text{\ and } m_1<-\frac{\beta}{\gamma+\beta}\big)
\text{\ or }
\big(m_1>\frac{1}{2}(m_0-\frac{\beta}{\gamma+\beta})\text{\ and } m_1<-\frac{\beta}{\gamma+\beta}\big)\bigg).
  \end{array}
\end{equation}
Then two  symmetric equilibrium points:
\begin{equation}
\label{Eq_point_13}
\begin{array}{cc}
    u^{1,3}_{\rm eq} = \pm (u^{x}_{\rm eq}, u^{y}_{\rm eq}, u^{z}_{\rm eq})=
    \pm \big(\tfrac{(\gamma+\beta)(m_{0}-m_{1})}{\gamma m_{1}+\beta m_{1}+\beta},
    \tfrac{\gamma(m_{0}-m_{1})}{\gamma m_{1}+\beta m_{1}+\beta},
    \tfrac{-\beta(m_{0}-m_{1})}{\gamma m_{1}+\beta m_{1}+\beta}\big),
\end{array}
\end{equation}
exist and the corresponding linearizations have the form:
\begin{equation}
\label{Matrix1_3}
    J(u^{1,3}_{\rm eq})=\left(
               \begin{array}{ccc}
                 -\alpha (m_{1}+1) & \alpha & 0 \\
                 1 & -1 & 1 \\
                 0 & -\beta & -\gamma \\
               \end{array}
             \right).
\end{equation}
For the zero equilibrium $u^{2}_{\rm eq} = \big(0,0,0\big)$
we have the following matrix of linearization
\begin{equation}
\label{Matrix2}
    J(u^{2}_{\rm eq})=\left(
               \begin{array}{ccc}
                 -\alpha (m_{0}+1) & \alpha & 0 \\
                 1 & -1 & 1 \\
                 0 & -\beta & -\gamma \\
               \end{array}
             \right).
\end{equation}
Remark that for $(m_0,m_1) \to (m_1,m_0)$
we have   $\big(J(u^{1,3}_{\rm eq}), J(u^{2}_{\rm eq})\big)
\to \big(J(u^{2}_{\rm eq}), J(u^{1,3}_{\rm eq})\big)$.
It means that the local bifurcations,
which occur at the symmetric equilibria $u^{1,3}_{\rm eq}$
and at the zero equilibrium $u^{2}_{\rm eq}$, are the same.
For the symmetric equilibria, the bifurcations occur,
when the parameter $m_{1}$ is varying,
for the zero equilibrium, when the parameter $m_{0}$ is varying.
The stability of equilibria depends on $m_{0}$ and $m_{1}$
and is determined by the eigenvalues
($\lambda_{1}$, $\lambda_{2}$, $\lambda_{3}$)
of the corresponding linearization matrices.

\begin{figure}[h]
\begin{center}
\includegraphics[width=0.7\textwidth]{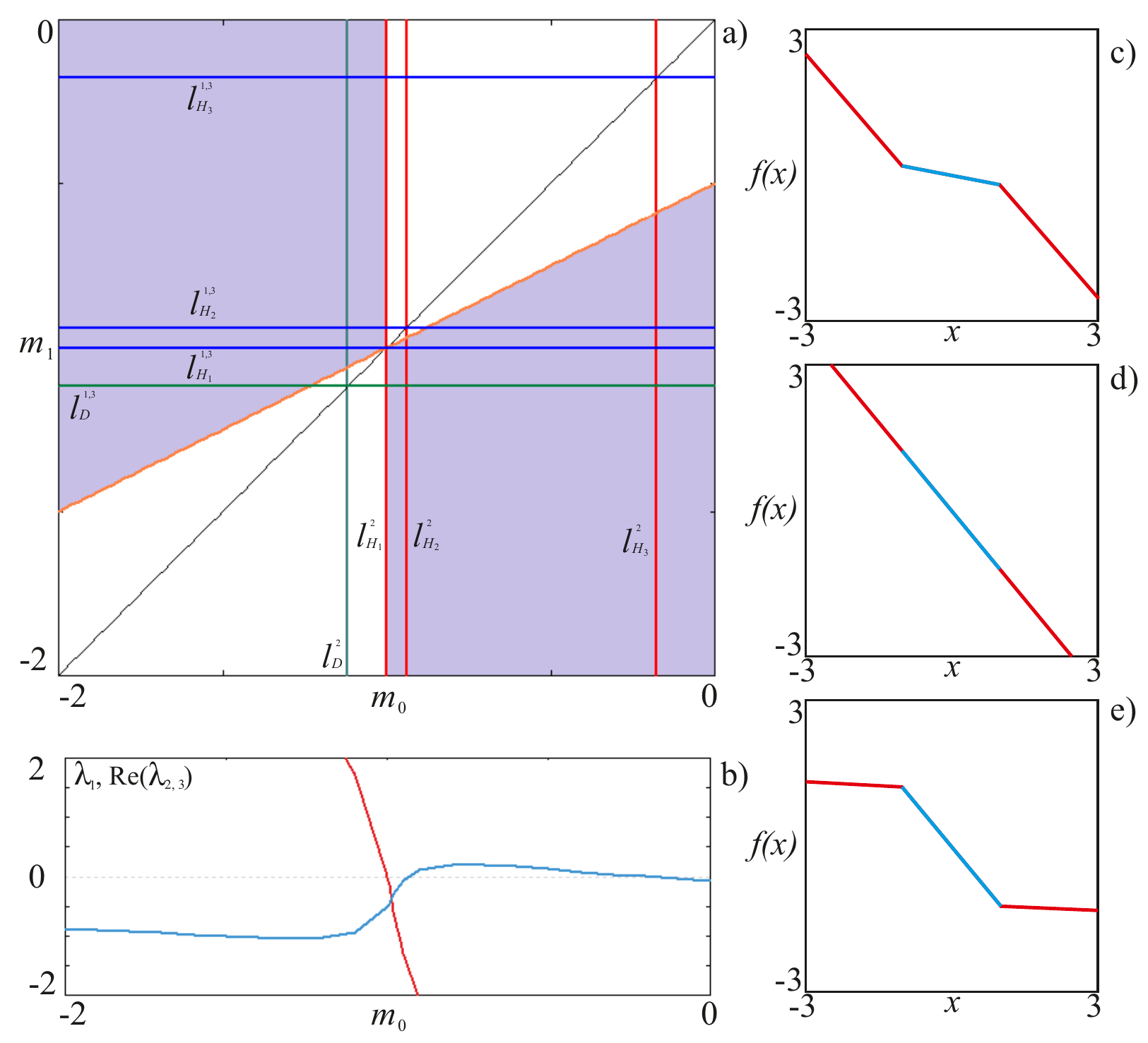}
\end{center}
\caption{a) Bifurcation lines of equilibrium points on
the parameter plane ($m_{0}$, $m_{1}$) for $\alpha=8.4$, $\beta=12$,
$\gamma=-0.005$, the red lines correspond to $u_{\rm eq}^2$
and the blue lines to $u_{\rm eq}^{1,3}$,
the areas of existence of the equilibria $u_{\rm eq}^{1,3}$
are filled by violet color;
b) the dependence of the real eigenvalue (red color) and
the real part of the complex
conjugate eigenvalues (blue color) on the parameter $m_{1}$; c) - e),
the examples of the voltage-current characteristics, c) $m_{0}=-0.2$, $m_{1}=-1.15$; d)
$m_{0}=-1.2$, $m_{1}=-1.2$; e) $m_{0}=-1.2$, $m_{1}=-0.05$.}
\label{Fig.1}
\end{figure}

Consider the following values of parameters
\begin{equation}\label{param}
  \alpha = 8.4, \beta = 12, \gamma = -0.005,
\end{equation}
which are close to the values, considered in \cite{LeonovKV-2011-PLA}
and are used for the construction of a hidden attractor.
Then for all equilibrium points,
one of the eigenvalues, $\lambda_{1}$ is always
real and can be positive or negative.
Two  other eigenvalues $\lambda_{2}$ and $\lambda_{3}$
are complex-conjugated and their real parts
can be also positive or negative.
Therefore we consider the following types of equilibria:
\begin{description}
  \item[-\ $F$] is a stable focus, $\lambda_{1}<0$, $Re(\lambda_{2,3})<0$;
  \item[-\ $SF$-$I$] is a saddle-focus of the first type:
  there are an unstable one-dimensional manifold and a stable two-dimensional manifold,
$\lambda_{1}>0$, $Re(\lambda_{2,3})<0$;
  \item[-\ $SF$-$II$] is a saddle-focus of the second type:
  there are a stable one-dimen\-sional manifold and an unstable two-dimensional manifold,
$\lambda_{1}<0$, $Re(\lambda_{2,3})>0$.
\end{description}

In Fig.~\ref{Fig.1}(a) is shown the plane of parameters $(m_0,m_1)$
with the bifurcation curves:
blue color denotes the bifurcation curves of the symmetric equilibrium stability,
red color denotes the bifurcation curves of the zero equilibrium stability.
Areas, filled by violet color, denote the areas of existence
of the symmetric equilibria $u_{\rm eq}^{1,3}$.
The domains filled by white color correspond to the existence of the only one equilibrium
(see conditions (\ref{lim})).
In Fig.~\ref{Fig.1}(b) are shown the plots of $\lambda_{1}$ (red color) and real part of $\lambda_{2,3}$
(blue color) versus the parameter $m_{0}$ for
the  equilibrium $u_{\rm eq}^{2}$,
where one can see the changes of the eigenvalues sign caused by a Hopf bifurcation.

In Fig.~\ref{Fig.1}(a) the following symbols are used for $u_{\rm eq}^{1,3}$ and $u_{\rm eq}^{2}$:
\begin{description}
  \item[-\ $l_{H_{1}}^{1,3}$ ($m_{1}\approx-1.0004$), $l_{H_{1}}^{2}$
($m_{0}\approx-1.0004$)] are the lines of the Hopf bifurcation
corresponding to the transition from the saddle-focus of the first type ($SF$-$I$)
to the stable focus ($F$);

\item[-\ $l_{H_{2}}^{1,3}$ ($m_{1}\approx-0.939$), $l_{H_{2}}^{2}$
($m_{0}\approx-0.939$)] are the lines of the Hopf bifurcation
corresponding to the transition from the stable focus ($F$)
to the saddle-focus of the second type ($SF$-$II$);

\item[-\ $l_{H_{3}}^{1,3}$ ($m_{1}\approx-0.1761$), $l_{H_{3}}^{2}$
($m_{0}\approx-0.1761$)] are the lines of the Hopf bifurcation
corresponding to transition from the saddle-focus of the second type ($SF$-$II$)
to the stable focus ($F$).

\end{description}

  As mentioned above, the parameters $m_{0}$, $m_{1}$ are characterized
by the slopes of  piecewise linear characteristic.
In Fig.~\ref{Fig.1}(c)-(e) are shown examples of voltage-current characteristic
for the Chua system \eqref{Chua} for different points of  the parameter plane.

\subsection{Numerical study of the parameter plane.
Bifurcation scenario of the hidden attractors transformations}
\label{section:2-2}
\begin{figure*}
  \centering
  \includegraphics[width=0.9\textwidth]{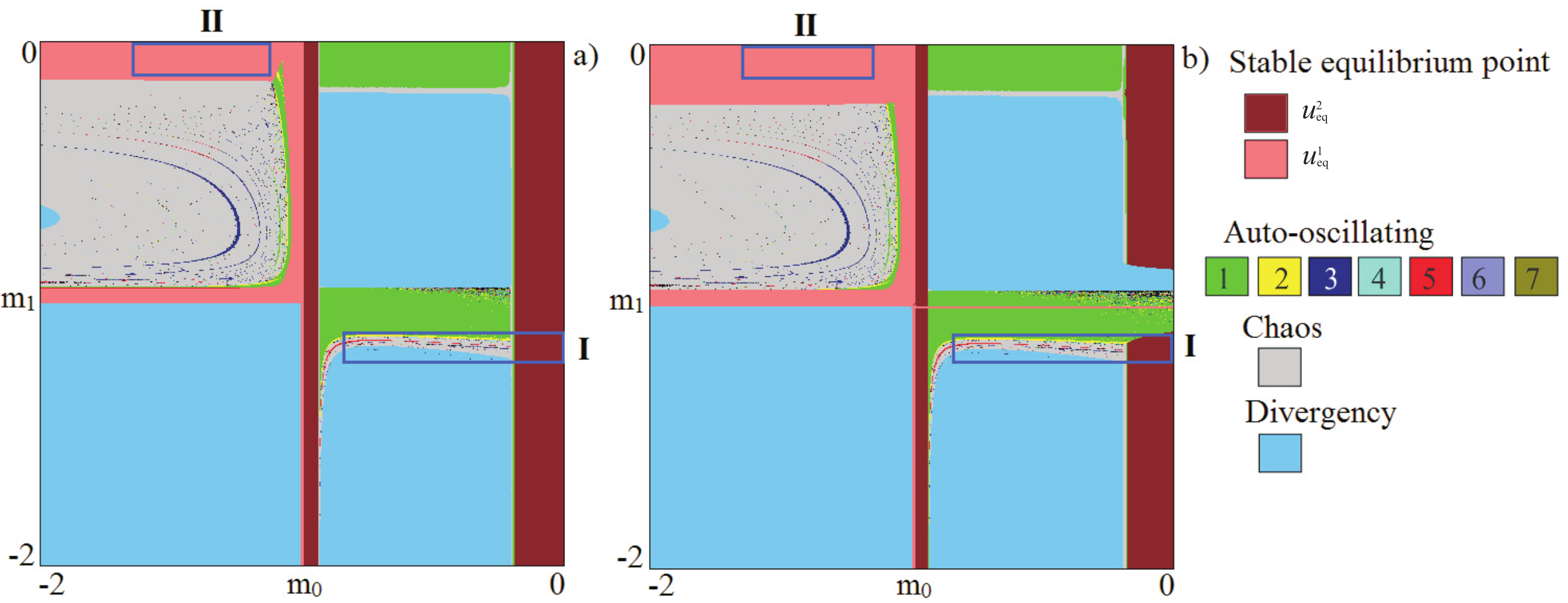}
\caption{Dynamics of the Chua system, $\alpha=8.4$, $\beta=12$, $\gamma=-0.005$,
and initial conditions: a) in the vicinity of the zero equilibrium
$u_{\rm eq}^{2}$, $x_{0}=y_{0}=z_{0}=0.0001$;
b) in the vicinity of one of the symmetric equilibrium points
$u_{\rm eq}^{1}$, $x_{0}=u_{\rm eq}^{x}+0.0001$,
$y_{0}=u_{\rm eq}^{y}+0.0001$, $z_{0}=u_{\rm eq}^{z}+0.0001$.}
\label{Fig.2}
\end{figure*}

Consider numerically the dynamics and the qualitative behavior of the Chua system (\ref{Chua})
in terms of parameters $m_{0}$, $m_{1}$.

In Fig.~\ref{Fig.2} the charts of dynamical regimes are shown on the parameter
plane ($m_{0}$, $m_{1}$). These charts are constructed in the following way.
The parameter plane ($m_{0}$, $m_{1}$)
is scanned with a small step. The dynamical regime, corresponding
to a point on the plane,
is determined according to the number of different points
in the Poincar\'{e} section,
defined by $z=0$ after a long enough transition process.
Initial conditions are the same for each value of parameters:
for the chart in Fig.~\ref{Fig.2}(a)
we take initial condition ($x_0$, $y_0$, $z_0$)=(0.0001, 0.0001, 0.0001)
in the vicinity of the zero equilibrium $u_{\rm eq}^{2}$.
For the chart in Fig.~\ref{Fig.2}(b)
we choose initial condition ($x_0$, $y_0$, $z_0$)=
($u_{\rm eq}^{x}$+0.0001, $u_{\rm eq}^{y}$+0.0001, $u_{\rm eq}^{z}$+0.0001)
in the vicinity of $u_{\rm eq}^{1}$ (one of the symmetric equilibria,
see (\ref{Eq_point_13})). Thus, we expect that the dynamical regimes,
which are visualized on these charts, are self-excited.
On the charts the  symmetric stable equilibrium points $u_{\rm eq}^{1,3}$
are marked by pink color, the zero stable equilibrium point $u_{\rm eq}^{2}$
 by maroon color,
the regime of divergency\footnote{Regime of divergency corresponds
to the regime, when the dynamical variables numerically grow  to infinity,
the detection of this regime is realized under the condition
$\sqrt{x^2+y^2+z^2}>10000$.} by blue color,
the chaotic dynamics\footnote{Chaotic regime is determined roughly:
if the number of discrete points
in the Poincar\'{e} section is more than 120.} by gray color.
The periodic oscillations with different periods are distinguished:
the green color for cycles of period-1,
the yellow color for cycles of period-2,
the dark-blue for cycles of period-3,
the blue color for cycles of period-4 and so on
(see the color legend in Fig.~\ref{Fig.2}).

We reveal that the complex dynamics of system (\ref{Chua})
is developed only in the case that three equilibria coexist.
Most of the areas where there is only one equilibrium
(see white domains in Fig.~\ref{Fig.1}),
belongs to the regime of divergency
(see the corresponding domains in Fig.~\ref{Fig.2}).
The exceptions are the bands, corresponding
to the stable zero equilibrium $u_{\rm eq}^{2}$ for
\[
  m_0>-0.1761 \text{\ and\ } m_1<\frac{1}{2}(m_0-\frac{\beta}{\beta+\gamma}),
\]
and the periodic oscillations, associated with
the Hopf bifurcation of the zero equilibrium $u_{\rm eq}^{2}$, for
\[
  m_1>-0.1761 \text{\ and\ } m_0>-\frac{\beta}{\beta+\gamma}.
\]

In the case of coexistence of three equilibria
the self-excited chaotic attractors are found (gray color).
However besides self-excited attractors,
here it is possible to find hidden attractors.

\begin{figure*}[ht]
  \center
  \includegraphics[width=0.9\textwidth]{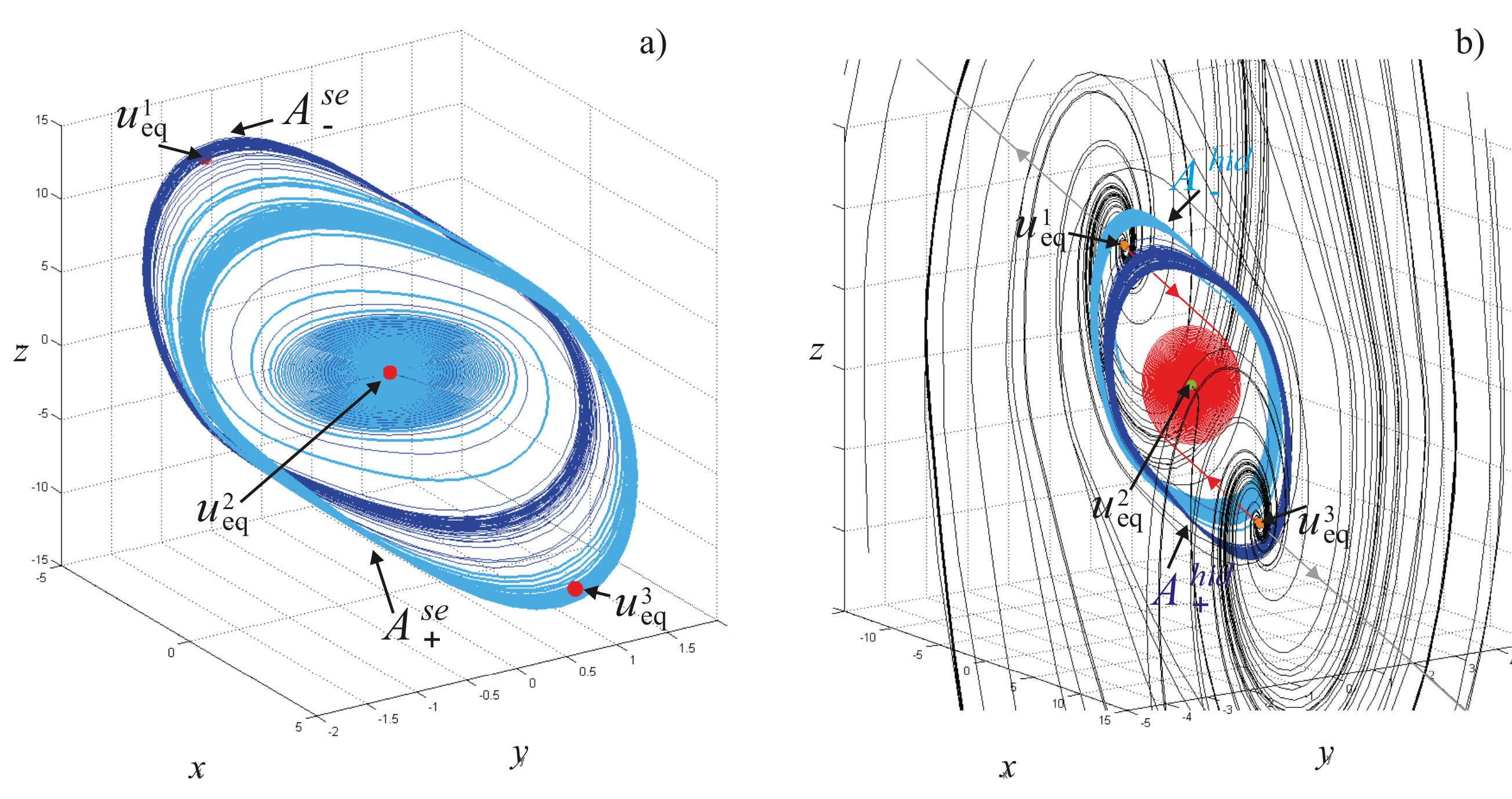}
  \caption{Self-excited and hidden attractors in the Chua system \eqref{Chua} with parameters
  $\alpha = 8.4$, $\beta = 12$, $\gamma = -0.005$:
  a) two symmetric self-excited Chua attractors (cyan and blue domains $A^{se}_{\pm}$)
  excited from unstable zero equilibrium
  (parameters $m_0 = -0.3$, $ m_1 = -1.12$);
  b) two symmetric hidden chaotic Chua attractors (cyan and blue domains $A^{hid}_{\pm}$).
  Red and gray trajectories from unstable manifold
  of the symmetric saddle-focuses equilibria $u_{\rm{eq}}^{1,3}$ (orange dots)
  are attracted to locally stable equilibrium $u_{\rm{eq}}^2$ (green dot)
  and infinity, respectively.
  Black trajectories are stable manifolds of $u_{\rm{eq}}^{1,3}$
  (parameters $m_0 = -0.121$, $m_1 = -1.143$).
  }
  \label{chua-se-ha-matlab}
\end{figure*}

In Fig.~\ref{Fig.2} the blue rectangle (\textbf{I})
is the area of the parameter plane where a hidden chaotic attractor
was discovered for the first time \cite{LeonovKV-2011-PLA}.
For $m_0<-0.1761$ (see, the line $l_{H_{3}}^{2}$
of the Hopf bifurcation for the zero equilibrium), all the observed attractors are
self-excited. In Fig.~\ref{chua-se-ha-matlab}(a) is shown an example of self-excited Chua attractor from this area of parameters. For $m_{0}>-0.1761$,
all dynamical regimes coexist with
stable zero equilibrium point. For $m_0>-0.1761$,
hidden attractors are observed
(an example of hidden Chua attractor is in Fig.~\ref{chua-se-ha-matlab}(b),
but in some small part of parameter plane a self-excited attractor is found:
the phase trajectories,
starting from a small neighborhood of the zero equilibrium $u_{\rm eq}^{2}$,
tend to the zero stable equilibrium,
but the phase trajectories,
starting from the vicinity of the symmetric equilibria $u_{\rm eq}^{1,3}$,
tend to an attractor (a limit cycle of period-1),
in which case the attractors are not hidden.
We consider this case in details in Section~3.1.

As mentioned in Section 2.1, the Chua system (\ref{Chua})
has symmetry with respect to the parameters $m_0$, $m_1$.
This means that the replacement ($m_0$, $m_1$) $\to$ ($m_1$, $m_0$)
in the Chua system \eqref{Chua} leads to
the replacement of stability of the equilibria: $u_{\rm eq}^{2}$ $\to$ $u_{\rm eq}^{1,3}$.
Thus, we can consider another area of possible existence of hidden attractors,
which is situated below the line $l_{H_{3}}^{1,3}$ in Fig.~\ref{Fig.1}(a),
i.e. before the Hopf bifurcation at the symmetric equilibria.
This area is denoted by the blue rectangle (\textbf{II}) in Fig.~\ref{Fig.2}.
For $m_1>-0.1761$ ($l_{H_{3}}^{1,3}$) there are hidden attractors
which cannot be visualized from the initial conditions in a small vicinity of the equilibria.

\section{Hidden twin attractors} \label{section:3}

\subsection{Merged twin attractors}
\label{section:3-1}
\begin{figure*}
  \centering
  \includegraphics[width=0.9\textwidth]{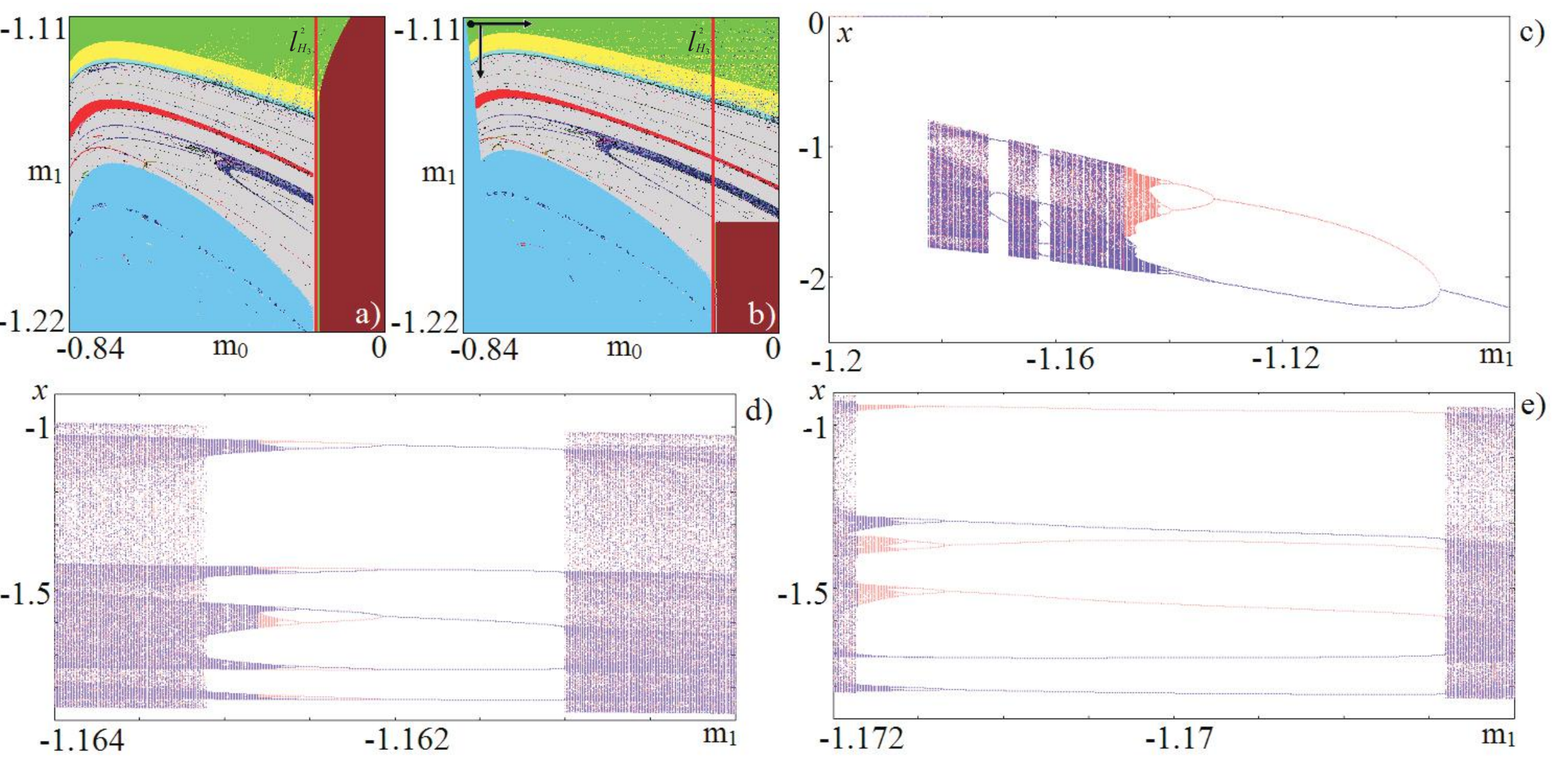}
\caption{a), b) magnified fragment (\textbf{I}) of the chart of
dynamical regimes for different initial conditions,
$\alpha=8.4$, $\beta=12$, $\gamma=-0.005$; c)
bifurcation diagram: red and violet colors correspond
to different initial conditions, $m_{0}=-0.121$; d), e)
magnified fragments of bifurcation diagram.}
\label{Fig.3}
\end{figure*}

Now we consider the dynamics of the Chua system \eqref{Chua}
with the parameters corresponding to the rectangle (\textbf{I}) in Fig.~\ref{Fig.2},
the zoom of which is shown in Fig.~\ref{Fig.3}(a).
For this area, for each point of parameter plane
the initial conditions are the same as in the vicinity of equilibrium $u_{\rm eq}^{1}$.
In Fig.~\ref{Fig.3}(b) the same fragment of the chart is constructed
by the so-called continuation method for choosing initial conditions, i.e.,
for each new value of parameter
as initial point, the same value is chosen as the final point
(obtained from the previous value of the parameter).
We use this method
to identify the area where hidden attractors exist.
In Fig.~\ref{Fig.3}(b) we mark a point in which we start our calculations,
and the arrows show the direction of scanning.

The analysis of stability of the equilibrium points in this area
(see Fig.~\ref{Fig.1}(a)) shows that for $m_0=0$ three equilibrium points exist:
two symmetric saddle-focus $SF$-$I$
and one stable focus $F$ (zero equilibrium).
In Fig.~\ref{Fig.3}(a) and (b) is shown
the bifurcation line of the loss of stability of the zero equilibrium
for $m_{0}\approx-0.1761$ ($l_{H_{3}}^{2}$).
In the area colored in the maroon,
the zero equilibrium point is characterized by one
negative real number and two  complex conjugate numbers with negative real
part; after crossing the bifurcation line ($l_{H_3}^2$)
the real parts of the conjugate-complex
eigenvalues become positive, and a stable focus is transformed
into a saddle-focus with a two-dimensional unstable manifold.
For the same parameters there exist also two symmetric equilibria,
which are characterized by one positive real number and
two complex-conjugate eigenvalues with a negative real part.

In the chart of dynamical regimes in Figs.~\ref{Fig.3}(a) and (b),
for fixed parameter
$m_{0}$ and decreasing parameter $m_{1}$,
one can observe a transition from a limit cycle of period-1 to
a chaotic attractor.
This transition corresponds to the Feigenbaum scenario
(the cascade of period-doubling bifurcations)
and it occurs in both the self-excited and the hidden attractors.

To analyze the birth of bifurcations, we construct
a bifurcation diagram versus parameter $m_{1}$.
In Fig.~\ref{Fig.3} are shown bifurcation diagram (c)
and its magnified fragments (d, e).
In Figs.~\ref{Fig.3}(c)-(e) is shown the dependence of the variable $x$, in the Poincar\'{e}
section by the plane $z=0$ (for $m_{0}=-0.1210$), on the parameter $m_{1}$.
In the diagram in Fig.~\ref{Fig.3}(c), we identify the value of the parameter $m_1^*\approx-1.1247$
corresponding to the transition from a self-excited attractor to a hidden attractor.
For $-1.0929<m_1<-1.0800$ the system exhibits one limit cycle with period-1
in Fig.~\ref{Fig.3}(c).
For $m_{1}\approx-1.0929$, the limit cycle is split into
two different period-1 limit cycles via a pitchfork bifurcation.
The pitchfork bifurcation is typical in the Chua system since
this system exhibits an inner symmetry. For $-1.1317<m_1<-1.0929$
two limit cycles of period-1 coexist, these cycles are
symmetric to each other. Notice that for $m_1\approx-1.1247$
these two period-1 limit cycles become hidden. For $m_1\approx-1.1317$
both limit cycles become limit cycles of period-2
via a period doubling bifurcation.
By continuous decreasing the parameter $m_1$,
after the sequence of period-doubling bifurcations,
as shown in Fig.~\ref{Fig.3}(c),
the two limit cycles are transformed into two different hidden chaotic attractors, respectively.
In this case these attractors coexist with a symmetric twin-attractor,
and a stable zero equilibrium point. For $m_{1}\approx-1.1483$
the twin-attractors are merged into one,
which by further decreasing the parameter $m_1$, forms an increasingly larger chaotic set.
For $m_1\approx-1.1609$, a periodic window of period-5 emerges
(in Fig.~\ref{Fig.3}(d) is shown a magnified fragment
near the periodic window of period-5).
The same scenario takes place for the period-5 cycle.
For $m_1\approx-1.1621$ the period-5 limit cycle is split into
two symmetric period-5 limit cycles via a pitchfork bifurcation.
For $-1.1626<m_1<-1.1621$
two limit cycles of period-5 coexist.
For $m_1\approx-1.1621$ both limit cycles bifurcate into two limit cycles of period-10
via a period-doubling bifurcation.
Upon further decreasing the parameter $m_1$, after a sequence of
period-doubling bifurcations, two hidden chaotic attractors emerge.
Then they merged at $m_1\approx-1.1628$, and collapse with
the chaotic set of the previous attractor at $m_1\approx-1.1631$.

Upon a further decrease of the parameter $m_1$,
a periodic window of period-3 emerges
(Fig.~\ref{Fig.3}(e) shows a magnified fragment
near a period-3 window).
In this case there is no pitchfork bifurcation, and
for $-1.1714<m_1<-1.1684$ two symmetric hidden cycles of period-3 coexist.
For $m_1\approx-1.1714$ both period-3 cycles become a pair period-6 limit cycles via a period doubling bifurcation.
Further decrease of the parameter $m_1$
gives rise to a cascades of period-doubling bifurcations. In this case
we can not see the merging of two chaotic attractors,
but for $m_1\approx-1.1719$ two hidden symmetric attractors
merge into a chaotic set.

\begin{figure}
\centering
 \includegraphics[width=0.7\textwidth]{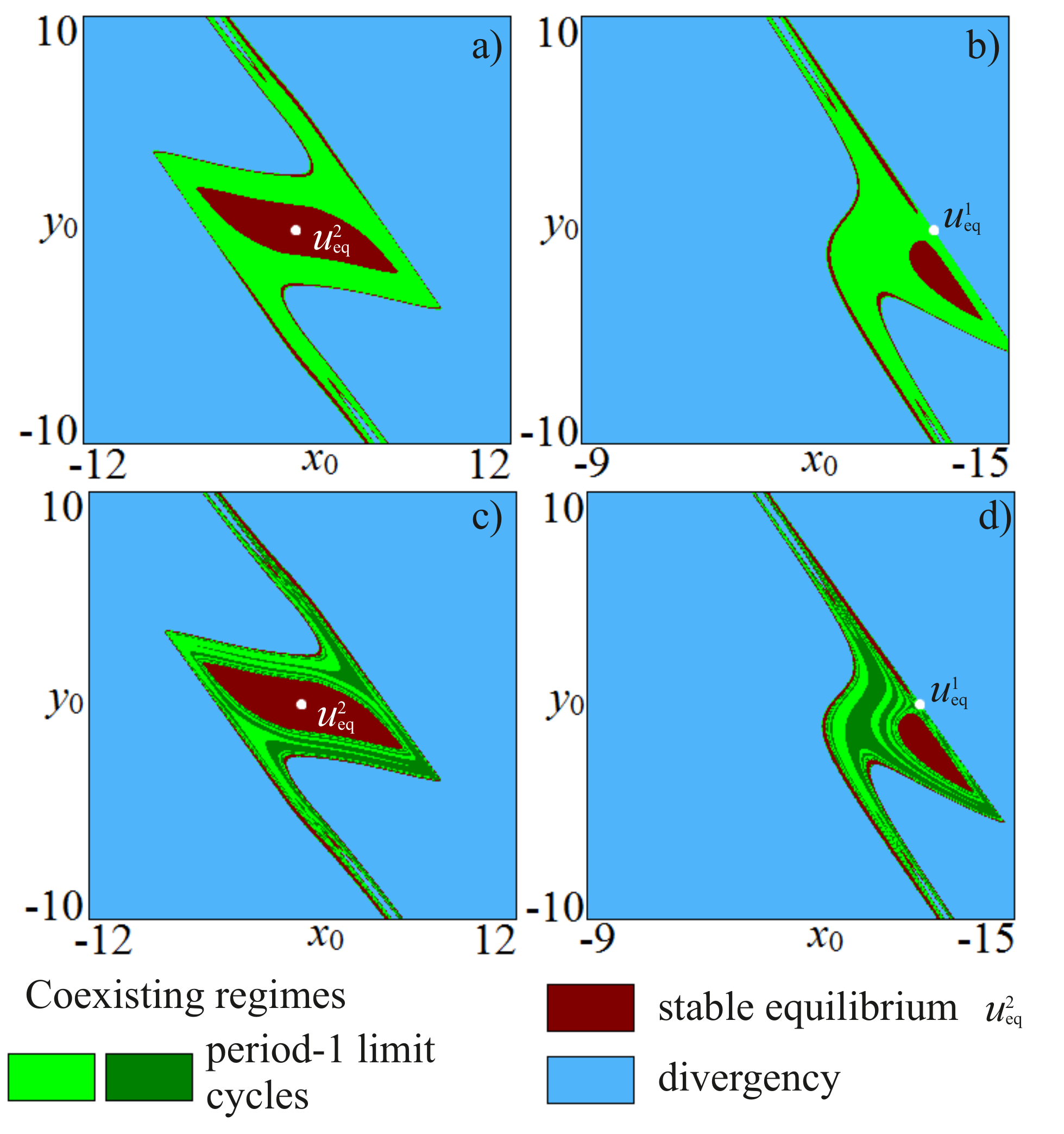}
\caption{Basins of attraction of coexisting self-excited attractors
of the Chua system with
$\alpha=8.4$, $\beta=12$, $\gamma=-0.005$, $m_{0}=-0.121$:
a) limit cycle of period-1 before pitchfork bifurcation $m_{1}=-1.09$,
  in the section defined by the plane $z_{0}=0.00001$;
b) in the section defined by the plane $z_{0}=u_{\rm eq}^z$;
c) period-1 limit cycles after pitchfork bifurcation
at $m_{1}=-1.1$, in the section defined by the plane $z_{0}=0.00001$;
d) in the section defined by the plane $z_{0}=u_{\rm eq}^z$.
}
\label{Fig.41}       
\end{figure}

\begin{figure}
  \centering
  \includegraphics[width=0.7\textwidth]{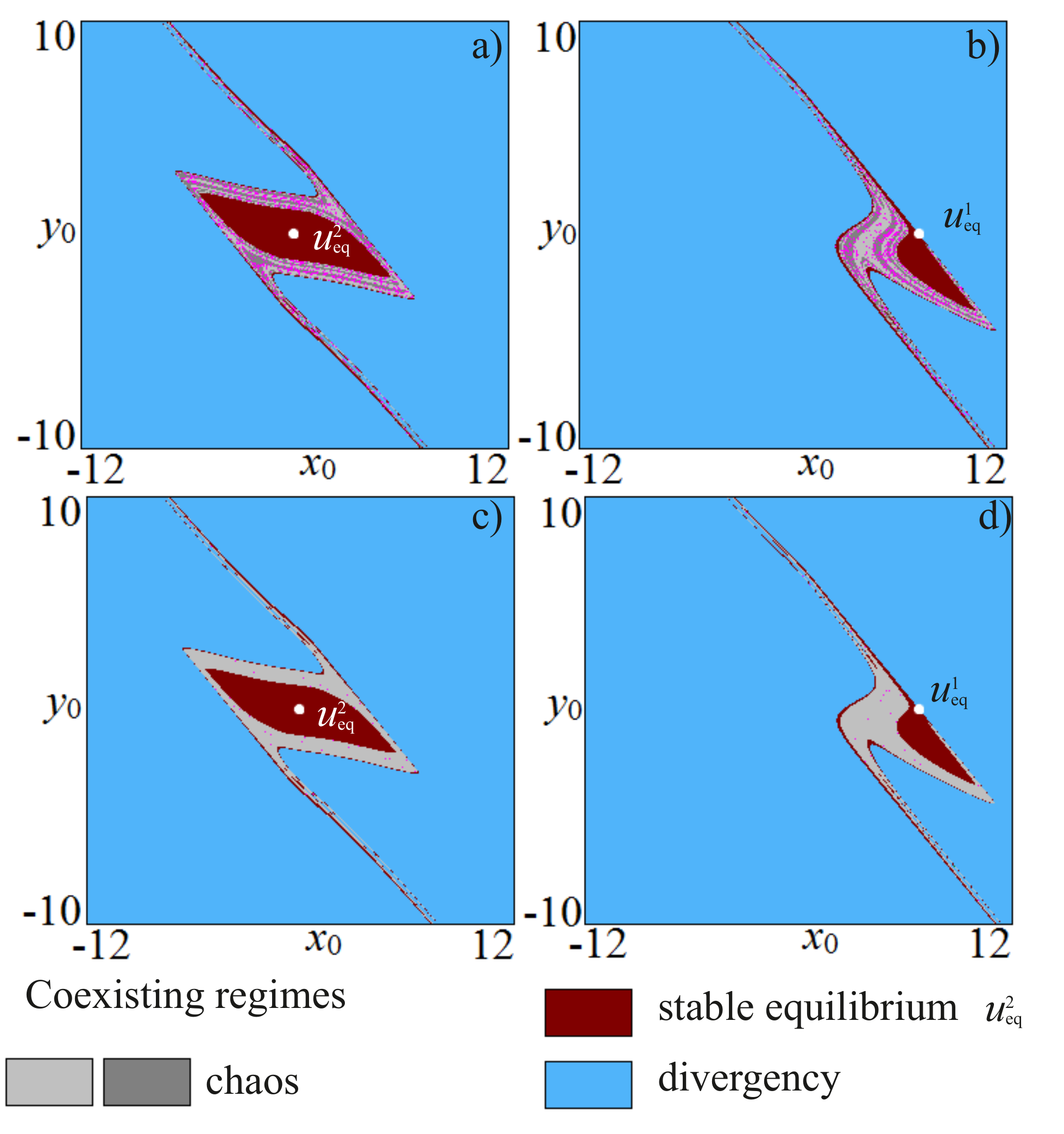}
\caption{Basins of attraction of coexisting
hidden attractors of the Chua system
with $\alpha=8.4$, $\beta=12$, $\gamma=-0.005$, $m_{0}=-0.121$:
a) twin hidden chaotic attractors for $m_{1}=-1.143$,
in the section defined by the plane $z_{0}=0.0001$;
b) in the section defined by the plane $z_{0}=u_{\rm eq}^z$;
c) merged hidden chaotic attractors for $m_{1}=-1.15$,
in the section defined by the plane $z_{0}=0.0001$;
d) in the section defined by the plane $z_{0}=u_{\rm eq}^z$.}
\label{Fig.42}       
\end{figure}

To analyze localization
of hidden attractors in the phase space, and transition from self-excited to hidden attractors,
we consider the basins of attraction under varying parameter.
Firstly, we consider the case that a self-excited attractor is realized:
$m_0=-0.121$, $m_1=-1.09$.
Since its basin of attraction in three-dimensional phase space
is difficult to analyze, we analyze
two-dimensional sections of this volume at various planes,
which correspond to two-dimensional planes of initial conditions.
To distinguish between hidden and self-excited attractors, the dynamical
behavior of the model in the vicinity of equilibria is crucial.
That is why we consider two sections of phase volume:
in the vicinity of the zero equilibrium $u_{\rm eq}^{2}$
and in the vicinity of one of the symmetric equilibria $u_{\rm eq}^{1}$
(for the other symmetric equilibrium $u_{\rm eq}^{3}$,
the structure of the basin is similar).

In Fig.~\ref{Fig.41} is shown the two-dimensional plane of initial
conditions for vicinities of equilibria points and for different values
of parameter $m_1$. The regime of divergency is marked by blue color,
the area of attraction of the stable zero equilibrium is marked by maroon color.
The areas of attraction regimes of different period-1 limit cycles are denoted
by two different green colors. The location of the equilibrium points
in the plane is identified by white dots. In Fig.~\ref{Fig.41}(a)
is shown a two-dimensional plane of initial conditions for fixed $z_{0}=0.0001$ in  the vicinity of the zero equilibrium $u_{\rm eq}^{2}$
(stable focus $F$).
For $m_1=-1.09$ in system (\ref{Chua}) the coexistence of a period-1 limit cycle,
before the pitchfork bifurcation, and the stable zero equilibrium is observed.
In Fig.~\ref{Fig.41}(a) is shown the structure of the basins of attraction of two
coexisting regimes. There is a rather large basin of attraction surrounding the stable zero equilibrium (maroon color),
and a large area of divergency.
Between these two areas we have an area of stable periodic oscillations,
which represents the basin of attraction of a period-1 limit cycle.
The boundary between the areas of divergency
and self-excited limit cycle
is indicated by a thick line, which corresponds
to the stable zero equilibrium.
In Fig.~\ref{Fig.41}(b) is shown a vicinity of one
of the symmetric points $z_{0}=u_{\rm eq}^z$ (saddle-focuses \textit{SF-I}).
The symmetric equilibrium state is located  on the boundary
between the basin of attraction of the stable limit cycle and the area of divergency.
In this case we cannot affirm that the limit cycle is a hidden attractor because
if we start to iterate a trajectory, with a randomly chosen initial state
in the vicinity of symmetric equilibrium points, it can either diverge from the initial state,
or tend to the limit cycle.

After the pitchfork bifurcation ($m_0=-1.1$) two symmetric limit cycles occur.
In Figs.~\ref{Fig.41}(c),(d) are shown two planes of initial conditions
with basins of attraction
of symmetric limit cycles in the vicinity of the two equilibria for
fixed third initial conditions $z_{0}=0.00001$ and $z_{0}=u_{\rm eq}^z$,
in which case, the two different green colors correspond to
the basins of attraction of the two symmetric limit
cycles ($m_{1}=-1.1$). In this case after the pitchfork bifurcation,
the basin of attraction of original limit cycle
is split into two basins of attraction of the two symmetric period-1 limit cycles.
These basins have a complex but symmetric structure.
In this case the equilibrium $u_{\rm eq}^{1}$
is also situated on the boundary of
the basins of attraction of the two limit cycles,
implying that the limit cycles are self-excited for $m_1>m_1^*$.

Next, we decrease parameter $m_1$ such that it becomes less then $m_1^*$.
In this case the attractors become hidden, while bifurcating into chaotic dynamics,
and we observe the corresponding changes in the structure in the plane of initial conditions.
In Fig.~\ref{Fig.3} is shown that hidden twin chaotic attractors exist,
for instance, at $m_{1}\approx-1.141$,
and with decreasing parameter $m_{1}$
these two attractors merged (for $m_{1}\approx-1.147$).
In Fig.~\ref{Fig.42} are shown the planes of initial conditions for the twin
chaotic attractors (a) and (b), and the merged chaotic attractor (c) and (d).
In Figs.~\ref{Fig.42}(a),(b) the basins of attraction of the two different
chaotic attractors are identified by different shaded gray colors.
The composition of the basins in the plane of initial states
near the vicinity of the zero equilibrium point is the same as
in the case of self-excited limit cycle.
We observe the basins of attraction of the twin chaotic attractors
and the basin of attraction of zero stable equilibrium point in the center.
But the structure of the plane of initial states in
the section near the vicinity of the symmetric equilibrium points
has a significant distinction.
The basin of attraction of the stable zero equilibrium $u_{\rm eq}^{2}$ at the center
is combined with the another part of the basin of attraction of the stable zero equilibrium point $u_{\rm eq}^{2}$
on the boundary of the area of divergency, and
a saddle equilibrium $u_{\rm eq}^{1}$ is located on the boundary
between the basin of attraction of the stable zero equilibrium point and the area of divergency.
Consequently, a twin chaotic attractor becomes hidden because
if we choose initial states near one of the equilibrium points,
then we cannot reach the chaotic attractors.

In Figs.~\ref{Fig.42}(c) and d are shown the same illustrations for merged
hidden chaotic attractor. In this case one can see that the basin
of attraction of the merged hidden attractor represents the combining
of the areas of attraction of each twin-hidden attractors.
In the vicinity of the saddle equilibrium we also see
only two possible regimes:
the stable zero equilibrium and the divergency.
It follows that the merged attractor is the hidden one.

\subsection{Separated twin-attractors}
\label{section:3-2}
\begin{figure*}
  \centering
  \includegraphics[width=0.9\textwidth]{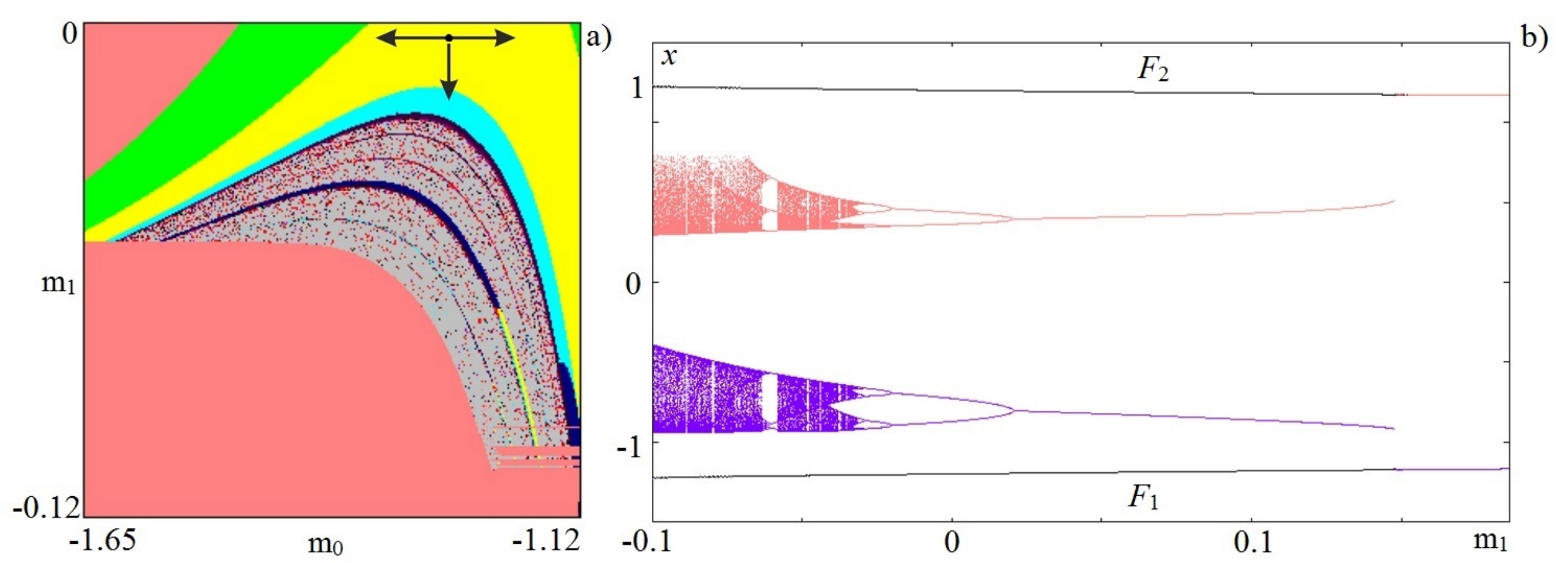}
\caption{a) the magnified fragment (\textbf{II})
of the chart of dynamical regimes for continuation method of changing initial conditions,
$\alpha=8.4$, $\beta=12$, $\gamma=-0.005$; b) bifurcations diagram:
black, red, and violet colors correspond to different initial conditions,
$m_{0}=-1.2$.}
\label{Fig.5}
\end{figure*}

Now we consider the dynamics of the Chua system (\ref{Chua})
and the features of hidden attractors in another area of the parameter plane,
which is marked by the blue rectangles (\textbf{II}) in Fig.~\ref{Fig.2}.
In Fig.~\ref{Fig.5}(a) is shown the zoom of fragment (\textbf{II}).
For the continuation method of changing initial conditions,
the starting point is denoted on the parameter plane,
and we scan the plane of parameters
in different directions in accordance with the arrows in this figure.

The analysis of stability of equilibria (Fig.~\ref{Fig.1}(a)) shows
that in this area there are two symmetric stable focuses ($F_1$, $F_2$)
and one saddle-focus of the first type (\textit{SF-I}) at the zero equilibrium point.

By numerical integration the trajectories starting from the vicinity of any equilibrium point
in rectangle (\textbf{II}) can reach only
one of the symmetric equilibrium points (Fig.~\ref{Fig.2}).
But we can see in Fig.~\ref{Fig.5}(a) that for some special initial conditions
it is possible to observe hidden attractors. In particular, the bifurcation scenario associated with the hidden attractors from the area of the parameter plane
(\textbf{II}) is the same as that in the area (\textbf{I}): one can observe
chaotic dynamics resulting from of a cascade of period-doubling bifurcations.

To analyze the bifurcations and transformations in this case,
let us consider a bifurcation diagram.
In Fig.~\ref{Fig.5}(b) are shown diagrams for different
initial conditions for $m_{0}=-1.2$, as a function of the parameter $m_1$ in the Poincar\'{e} section by the plane $z=0$.
The black lines correspond to
initial condition near the symmetric equilibrium points (the scanning of
parameter $m_{1}$ was realized by the continuation method to choose
initial conditions), and these lines mark the coexisting stable
focuses ($F_{1}$ and $F_{2}$). For the red and violet bifurcation diagrams
the initial conditions are chosen in the following way: $x_{0}=\mp1.2$,
$y_{0}=\mp0.0005$, $z_{0}=0$, respectively. For these initial conditions
for $m_{1}=0$ we have two symmetric period-2 limit cycles,
and from these points we scan the parameter intervals $m_{1}$ $[-0.2, 0.1]$
in two directions (to the right and to the left).

\begin{figure}
  \centering
  \includegraphics[width=0.7\textwidth]{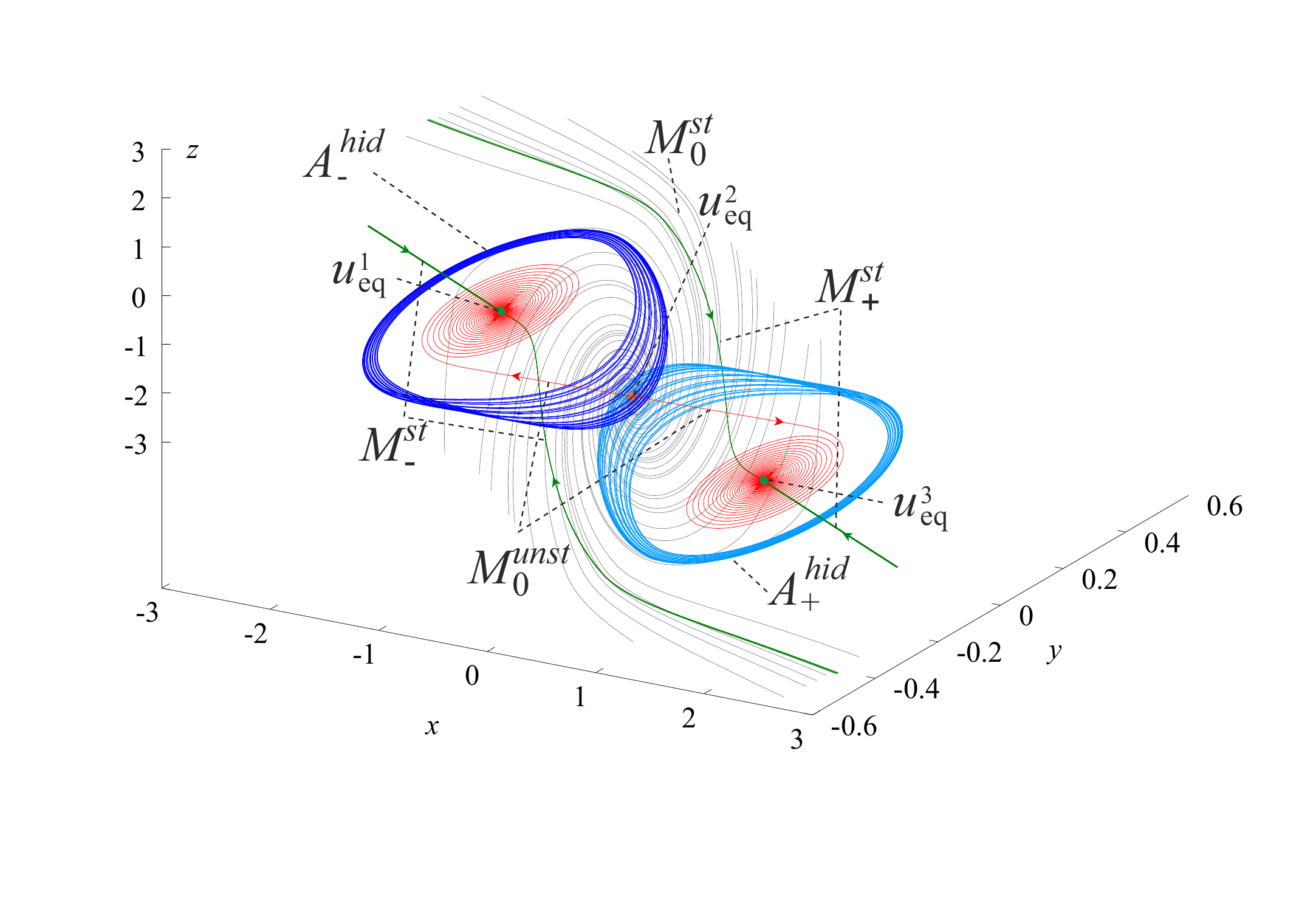}
\caption{Example of two symmetric hidden chaotic Chua attractors
    (blue and cyan domains $\mathcal{A}^{\rm hid}_{\pm}$), $m_1 = -0.05$.
  Red trajectories from unstable manifold ${\rm M_{0}^{\rm unst}}$
  of the zero saddle-focus equilibrium $u_{\rm{eq}}^2$ (orange dot)
  are attracted to locally stable equilibria $u_{\rm{eq}}^{1,3}$ (green dots);
  gray trajectories are stable manifold ${\rm M_{0}^{\rm st}}$
  of $u_{\rm{eq}}^{2}$;
  green trajectories are stable manifolds ${\rm M_{\pm}^{\rm st}}$
  of $u_{\rm{eq}}^{1,3}$.}
\label{Fig.7nnn}
\end{figure}

For $m_{1}=0.2$ there are two  symmetric stable equilibrium points.
If the parameter $m_{1}$ decreases, then at $m_{1}\approx0.15$
a limit cycle emerges near each  symmetric equilibrium point.
In the bifurcation diagram one can see the hard birth of the cycles,  because of the form of nonlinearity
(piecewise-linear characteristic of the Chua circuit).
 As $m_1$ decreases, we see stable focuses and two coexisting limit cycles,
undergoing a period-doubling bifurcation and transition to chaos.
However, for this area of parameter plane there
is no a pitchfork bifurcation of limit cycles
and a merging of chaotic attractors.
In this case two bifurcation diagrams in Fig.~\ref{Fig.5}~(b)
do not cross each other and are separated by the saddle point at the zero equilibrium. In Fig.~\ref{Fig.7nnn} the example of two symmetric hidden chaotic Chua attractors are shown for $m_1=-0.05$. By red, gray and green colors in Fig.~\ref{Fig.7nnn} are shown
the stable and unstable manifolds in the vicinity of equilibrium points.
Gray and green trajectories were constructed by integration in inverse time
for initial conditions in the vicinity of
equilibrium points (gray lines are near saddle-focus,  green lines are near stable focuses).
Red trajectories tend to the symmetric equilibria
and are obtained by the integration in forward time
near the zero equilibrium.

\begin{figure}
  \centering
  \includegraphics[width=0.9\textwidth]{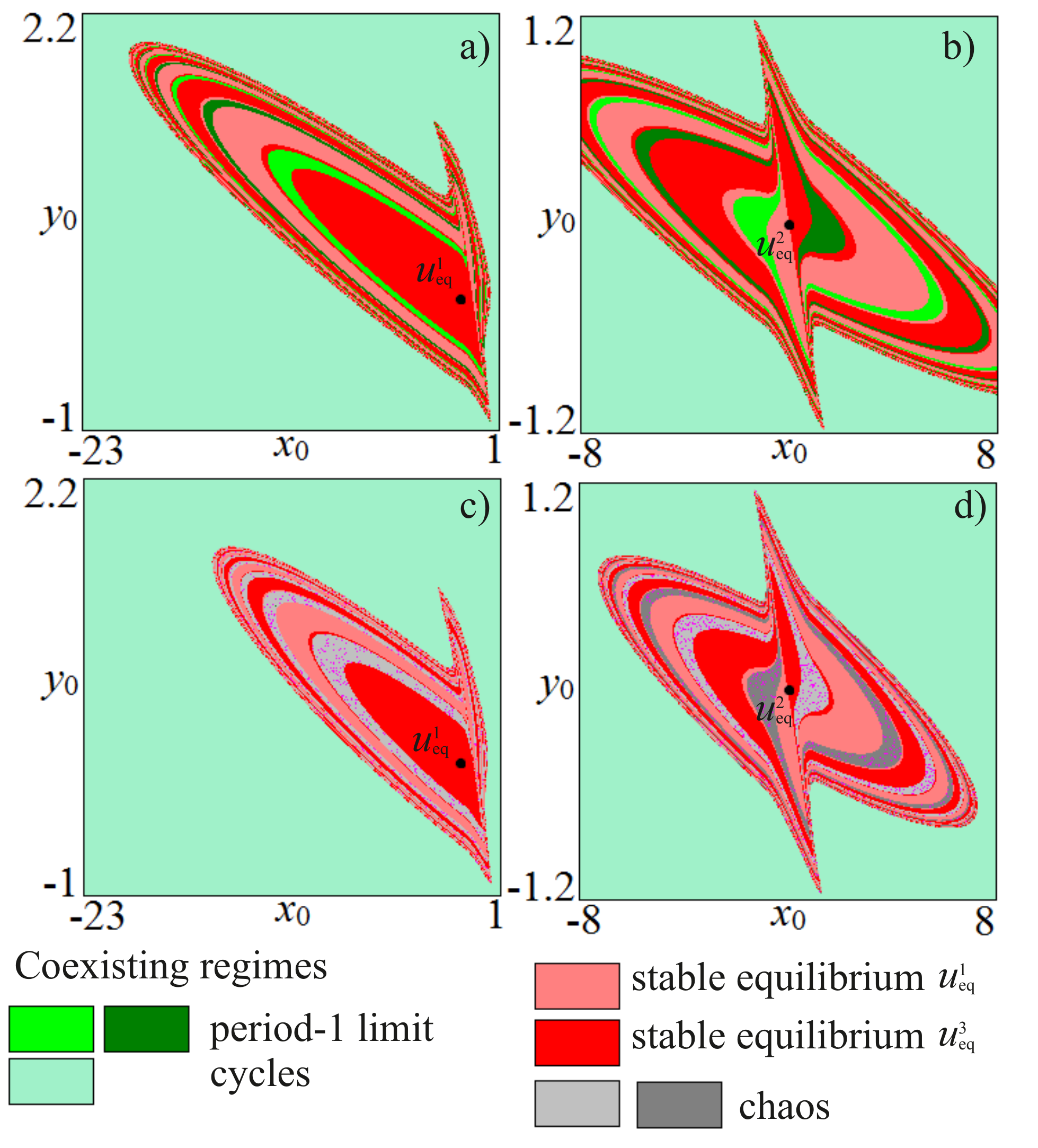}
\caption{Basins of attraction of coexisting attractors of the Chua system
\eqref{Chua} $\alpha=8.4$, $\beta=12$, $\gamma=-0.005$, $m_{0}=-1.2$:
a) twin hidden period-1 attractors with $m_{1}=0.1$, and cross-section by the plane $z_0=u_{\rm eq}^z$;
b) cross-section by the plane $z_0=0.0001$; c) twin hidden chaotic attractors with $m_{1}=-0.05$,
cross-section by the plane $z_0=u_{\rm eq}^z$; d) cross-section by the plane $z_0=0.0001$.}
\label{Fig.6}       
\end{figure}

\begin{figure*}[!ht]
  \centering
\includegraphics[width=0.9\textwidth]{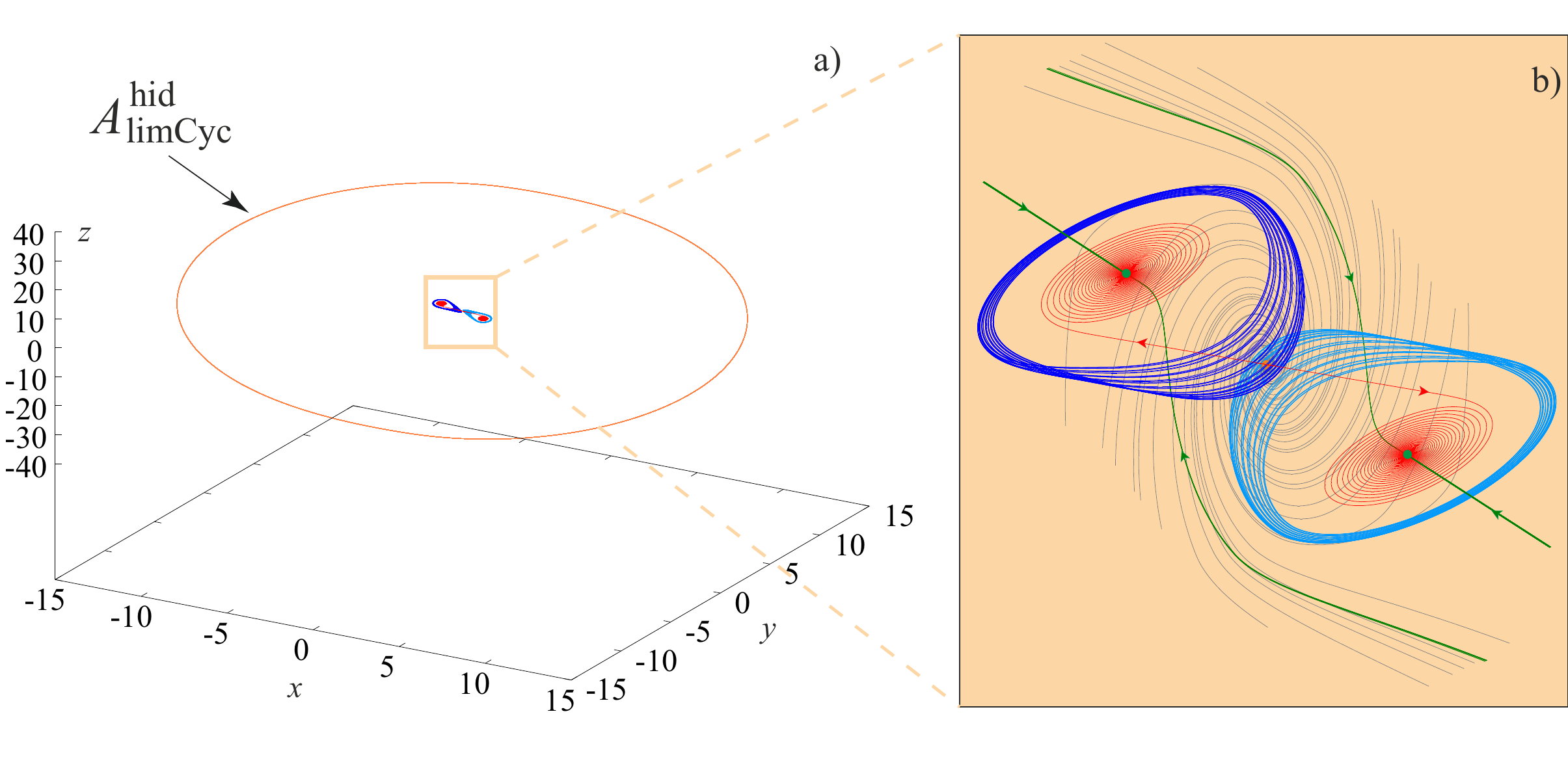}
  \caption{
    Multistability with 5 coexisting attractors in the Chua system (\ref{Chua})
    with $\alpha = 8.4$, $\beta = 12$, $\gamma = -0.005$,
    $m_0 = -1.2$, $ m_1 = -0.05$.
    Coexistence of  hidden periodic attractor
   (orange trajectory $\mathcal{A}^{\rm hid}_{\rm limCyc}$
    is a stable limit cycle)
    and two symmetric hidden chaotic Chua attractors
    (blue and cyan domains).
  Here the basins of attraction of the periodic and the two symmetric Chua attractors
  do not intersect with a small neighborhood
  of the equilibria, thus, the attractors are hidden.
  }
  \label{fig:hidden02}
\end{figure*}

To analyze the structure of the phase space and the localization
of hidden attractors in the phase space, we consider
two-dimensional planes of initial conditions.
Then we can study the basins of attraction
of co-existing attractors. As in the previous case
we consider two-dimensional sections of the
three-dimensional phase space of initial states
in the vicinity of one of the stable symmetric equilibria
$u_{\rm eq}^{1}$, and in the vicinity
of the zero equilibrium $u_{\rm eq}^{2}$.
In Fig.~\ref{Fig.6} is shown the structure of areas of attraction
for different parameters of $m_{1}$ and $m_{0}=-1.2$,
and for different cross-section of the phase space.
In Fig.~\ref{Fig.6}(a) and (b) are shown the basins of attraction of coexisting
attractors for $m_{1}=0.1$ in the vicinity of one stable focus $F_1$ (a)
and in the vicinity of the zero saddle equilibrium (b),
the location of equilibrium points are marked in the plane by black dots.
In this case in the bifurcation diagram one can see two coexisting symmetric equilibrium points and two symmetric cycles of period-1.
Also it is observed a new period-1 limit cycle, surrounding all regimes described above,
and the dynamics near the equilibria is developed inside this limit cycle
of sufficiently large radius. So, for this area of parameters
we have five coexisting attractors.
We shaded the areas of attraction of different symmetric period-1 limit cycles
by green color. The area of attraction of the outside
limit cycle is marked on the plane by light green color.
We use the pink and red colors to denote the basins of attraction of
the two symmetric equilibria $u_{\rm eq}^{1}$ and $u_{\rm eq}^{3}$, respectively.

Firstly, we consider a vicinity of the stable equilibrium (Fig.~\ref{Fig.6}(a)).
In the vicinity of the stable equilibrium one can see a basin of attraction
of one of the symmetric stable equilibrium points.
Also, there are the basin of attraction of another symmetric stable equilibrium point,
and the symmetric basins of attraction of two symmetric limit cycles.
The complex structure of their basins is represented by the area of attractions
in the form of bands,
which are spiralled together,
and their boundaries have self-similar patterns, i.e., fractal structures.
Also, there is a basin of attraction of the external limit cycle,
which surround all other basins of attraction.

Then we consider a plane of initial conditions and the basins of attraction
of different attractors in the vicinity of the saddle equilibrium point
(Fig.~\ref{Fig.6}(b)). We can see that the phase trajectories, starting
from the vicinity of the saddle point, can reach
one of the stable symmetric equilibria only. The zero equilibrium point $u_{eq}^2$
is located on a boundary between the attracting areas of
different symmetric stable equilibria.
The areas of attraction are symmetric to each other,
and a boundary between these areas represents the stable manifold
of the zero saddle-focus. Consequently, if we choose initial conditions in the vicinity
of any equilibrium point, we will reach one of the stable equilibrium points
and, thus, all of the limit cycles are hidden attractors.

Let us decrease the parameter $m_1$ so that chaotic dynamics emerges, and consider the basins of attraction of the coexisting stable symmetric equilibria,
the chaotic attractors, and the external limit cycle. In Fig.~\ref{Fig.6}(c)
and (d) are shown two planes of initial conditions in the vicinity
of the equilibrium points for $m_1=-0.05$.
By two shades of gray color we identify the basins of attraction
of the coexisting twin-chaotic attractors.
It is rather easy to distinguish the basins of attraction
for these chaotic attractors because the attractors
in the phase space are separated by the zero saddle equilibrium. For decreasing parameter $m_{1}$ the structure of the basin of attraction
persists. In place of the areas of two symmetric period-1 limit cycles
one sees the basins of attraction of the symmetric chaotic attractors.
In this case the order of alternation of the basin of attraction
of different attractors remains the same.
The structure of the basin of attraction
in the vicinity of the saddle equilibrium point persists:
the saddle point
is located on the boundary of the basins of attraction of
two  symmetric equilibrium points, and the boundary corresponds
to an stable manifold of the saddle point.

In Fig.~\ref{fig:hidden02}~(a),(b) is shown a structure of
the phase space in the above case,
where we see a coexisting large stable limit cycle (orange color in Fig.~\ref{fig:hidden02}~(a))
and two separated hidden chaotic Chua attractors (blue and cyan domains
Fig.~\ref{fig:hidden02}~ (b))
from Fig.~\ref{Fig.7nnn}.
The basins of attraction of periodic and chaotic attractors
do not intersect with a small neighborhood
of the equilibria, thus, the attractors are hidden.
Therefore in this case there are 5 coexisting attractors:
two stable equilibria, one hidden limit cycle, and two hidden ``twin'' attractors.

Thus, we reveal two area on the parameter plane ($m_0$, $m_1$),
where we observe hidden attractors.
Observe that for the physical realization of the Chua circuit
and observation of hidden attractors
we need nonnegative parameters $\alpha$, $\beta$, $\gamma$.
For example, both configurations of hidden attractors cited above are
observed for the case $\gamma=0$ in \cite{RochaM-2015,RochaM-2016-HA};
and for the positive $\gamma$ one may consider, for example, the following two sets of parameters:
$\alpha=8.4562, \beta = 12.0732, \gamma = 0.0052, m_0 =-0.1768, m_1= -1.1468$ and
$\alpha=8.4, \beta = 12, \gamma = 0.005, m_0 = -0.12, m_1 = -1.143 $.

Note that the existence of hidden attractors in the Chua system
can be effectively predicted by the describing function method (DFM) \cite{LeonovKV-2011-PLA,RochaM-2015,KuznetsovKLMS-2017-IFAC}.
The classical DFM (see, e.g. \cite{KrylovB-1947,Khalil-2002})
is only an approximate method which gives the information on
the frequency and amplitude of periodic orbits.
However DFM may lead to wrong conclusions\footnote{
Well-known Aizerman's and Kalman's conjectures
on the absolute stability of nonlinear control systems
are valid from the standpoint of DFM
which may explain why these conjectures were put forward.
Nowadays, various counterexamples to these conjectures
(nonlinear systems, where the only equilibrium, which is stable,
coexists with a hidden periodic oscillation) are known
(see, e.g. \cite{Pliss-1958,Fitts-1966,Barabanov-1988,BernatL-1996,LeonovBK-2010-DAN,LeonovK-2011-DAN}
and surveys \cite{BraginVKL-2011,LeonovK-2013-IJBC};
the corresponding discrete examples are considered in \cite{Alli-Oke-2012-cu,HeathCS-2015}).
}
about the existence of periodic orbits
and does not provide initial data for the localization of periodic orbits.
But for the systems of special type with a small parameter, DFM can be rigorously justified.
For this purpose, following references \cite{LeonovKV-2011-PLA,LeonovK-2013-IJBC},
we introduce a coefficient $k$ and represent the linear part and nonlinearity in \eqref{Chua-2} as follows:
\begin{equation} \label{ChuaLphi}
\begin{aligned}
	&{P_0} ={P}+k{qr}^*= \left(
     \begin{array}{ccc}
       -\alpha(m_1+1+k)& \alpha & 0 \\
       1 & -1 & 1 \\
       0 & -\beta & -\gamma \\
     \end{array}
   \right), \\
   & \psi(\sigma) = \phi(\sigma) - k\sigma = (m_0-m_1) \, {\rm sat}(\sigma) - k\sigma,
\end{aligned}
\end{equation}
where $\lambda^{{P}_0}_{1,2} = \pm i\omega_0$, $\lambda^{{P}_0}_{3} = -d < 0$.
Then we consider a small parameter $\varepsilon$, change $\psi(\cdot)$ by $\varepsilon\psi(\cdot)$,
and reduce by a non-singular linear transformation ${w} = {Su}$
system \eqref{ChuaLphi} to the following form
\begin{equation} \label{ChuaDiag}
  \begin{aligned}
  &
  \dot w = {A}{w} + {b}\varepsilon\psi({u}^*{y}),
  \\ &
   {A} =
   \left(
    \begin{array}{ccc}
       0 & -\omega_0 & 0 \\
       \omega_0 & 0 & 0 \\
       0 & 0 & -d \\
     \end{array}
   \right),
   \
   {b} = \left( \begin{array}{c} b_1 \\ b_2 \\ b_3 \\ \end{array}\right),
   \
   {c} = \left( \begin{array}{c} 1 \\ 0\\ -h \\ \end{array} \right).
  \end{aligned}
\end{equation}

{\bf Theorem} \cite{LeonovKV-2011-PLA,LeonovK-2013-IJBC,KuznetsovKLMS-2017-IFAC}
{\it Consider the describing function
$
  \Phi(a)=\int_0^{2\pi\!/\omega_0}\psi(a\,\cos(\omega_0t))\cos(\omega_0t)dt.
$
If there exists a positive $a_0$ such that}
$
  \Phi(a_0)=0, \quad b_1\Phi'(a_0) < 0,
$
{\it then system \eqref{ChuaDiag} has a stable\footnote{See detailed discussion in \cite{LeonovK-2013-IJBC}.} periodic solution with the initial data
$w_0=\big(a_0 + O(\varepsilon), 0, O(\varepsilon)\big)$
and period $T = \frac{2 \pi}{\omega_0} + O(\varepsilon)$.
}

This theorem gives an initial point
for the numerical computation of periodic solution (starting attractor)
in the system with small parameter.
Then, using the method of numerical continuation and gradually increasing $\varepsilon$,
one can numerically follow the transformation of the starting attractor.

It turns out that for the numerical localization of the considered hidden attractors
we can skip the multistep procedure based on numerical continuation
and use the initial data $u_0=S^{-1}w_0$
for the localization of hidden attractors in the initial system ().
For the parameters
$
	\alpha = 8.4, \beta = 12, \gamma = -0.005, m_0 = -1.2, m_1 = -0.05
$
we get:
a)
$k = -0.8890, \omega_0 = 2.0260, a_0 = 1.5187$
and the corresponding initial data
$\pm(1.5187, 0.0926, -2.1682)$
allows us to visualize two symmetric hidden chaotic attractors;
b) $k = -0.1244, \omega_0 = 3.2396, a_0 = 11.7546$
and the corresponding initial data $(11.7546, 9.7044, -16.7367)$
allows us to localize the hidden periodic attractor (see Fig.~\ref{fig:hidden02}).
For the parameters
$
	\alpha = 8.4, \beta = 12, \gamma = -0.005,
     m_0 = -0.121, m_1 = -1.143
$
we get $k=0.2040, \omega_0 = 2.0260, a_0=6.3526$
and the corresponding initial data
$\pm(6.3526,    0.3874,   -9.0694)$
allows us to visualize two symmetric hidden chaotic attractors
(see Fig.~\ref{chua-se-ha-matlab}(b)).

\section{Scenario of the birth of hidden attractors}

In Sections 3.1 and 3.2 we have shown the opportunity of existence of hidden attractors in different areas of the parameter plane. In order to study the scenario of the emergence of hidden attractors
we use numerical bifurcation analysis by the software package XPP AUTO \cite{ErmentroutXPPAUTO-2002}.

\subsection{Formation of separated hidden attractors}
\begin{figure*}
  \centering
  \includegraphics[width=0.9\textwidth]{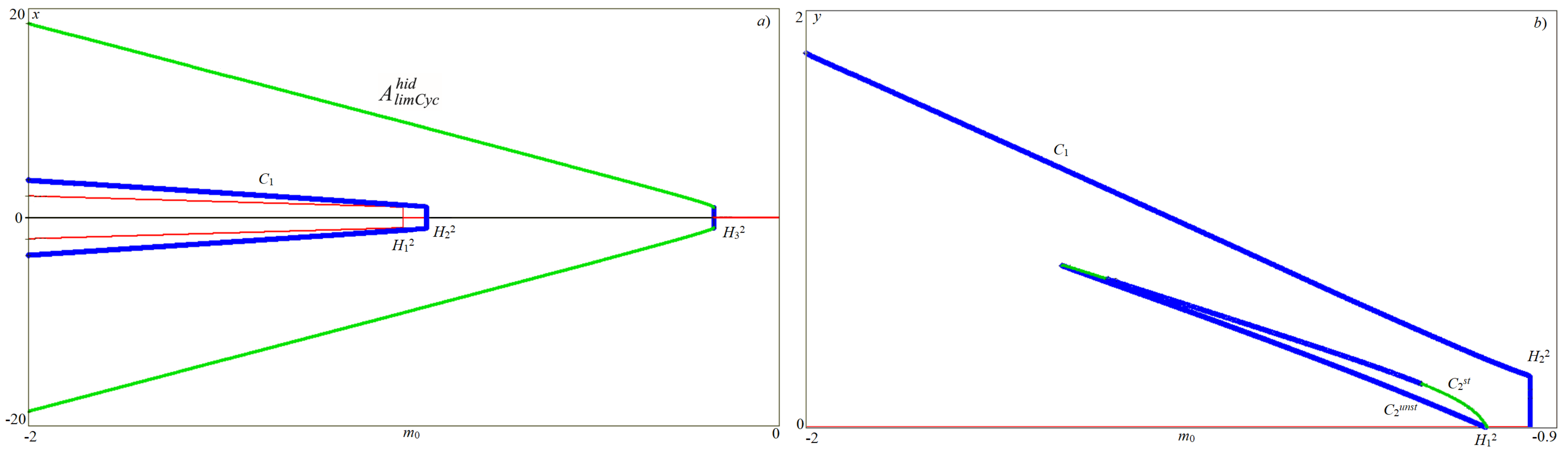}
\caption{Bifurcation diagram of the Chua system (\ref{Chua}), $\alpha=8.4$, $\beta=12$, $\gamma=-0.005$, $m_1=-0.05$.}
\label{Fig.10}
\end{figure*}

Firstly we consider hidden attractors from the area (\textbf{II}) in the parameter plane ($m_0$, $m_1$) (Fig.~\ref{Fig.2}). In Fig.~\ref{Fig.10}(a) is shown the bifurcation diagram of the Chua system (\ref{Chua}) for parameters (\ref{param}) and $m_1=-0.05$.
In the diagram red and black color denote stable and unstable equilibrium points, green and blue colors denote stable and unstable limit cycles, respectively.
At $m_0\approx-0.1761$ the Hopf bifurcation ($H_3^2$) takes place, that is in a good agreement with the results obtained by linear analysis in Section 2.1 and by numerical simulations in Section 3.2. In this case the supercritical Hopf bifurcation occurs: the zero equilibrium point loses stability and a limit cycle $A_{limCyc}^{hid}$ is born. With decreasing of parameter $m_0$ the radius of limit cycle is increased. At $m_0 \approx -0.939$ a second Hopf bifurcation ($H_2^2$) emerges,
where the zero equilibrium becomes stable, which is accompanied by the hard birth of an unstable limit cycle $C_1$.
The radius of the internal unstable cycle $C_1$ surrounding the zero equilibrium increases initially according to square root of 2, upon a further decrease of the parameter $m_0$.
At $m_0 \approx -1.0004$ occurs a third Hopf bifurcation ($H_1^2$), which is a pitchfork bifurcation of the zero equilibrium, in which case two stable symmetric equilibria are born, and the zero equilibrium become unstable.
In this case the limit cycle $C_1$ surrounds all equilibria, splits the limit cycle $A_{limCyc}^{hid}$ and equilibria in the phase space, and forms the boundary of the basin of attraction of the limit cycle $A_{limCyc}^{hid}$. The pitchfork bifurcation ($H_1^2$) is accompanied by the occurrence of two symmetric pairs of limit cycles $C_{2}^{st}$ and $C_{2}^{unst}$,
which are denoted in the magnified fragment of the diagram in Fig.~\ref{Fig.10}(b).
The birth of the limit cycles is a result of a saddle-node bifurcation,
i.e. a pair of limit cycles (stable and unstable) are born,
along with its identical symmetric pair.
Thus, the stable limit cycle $C_{2}^{st}$ is surrounded by the unstable limit cycle $C_1$ on the one side, and by the unstable limit cycle $C_{2}^{unst}$ on the other side,
making it unreachable from the vicinity of stable equilibria points and also from the vicinity of unstable equilibrium points. Thus, the limit cycles $C_2^{st}$, and the chaotic attractor, which occur on the base of this limit cycle are hidden.

\subsection{Formation of merged hidden attractors}
\begin{figure*}
  \centering
  \includegraphics[width=0.9\textwidth]{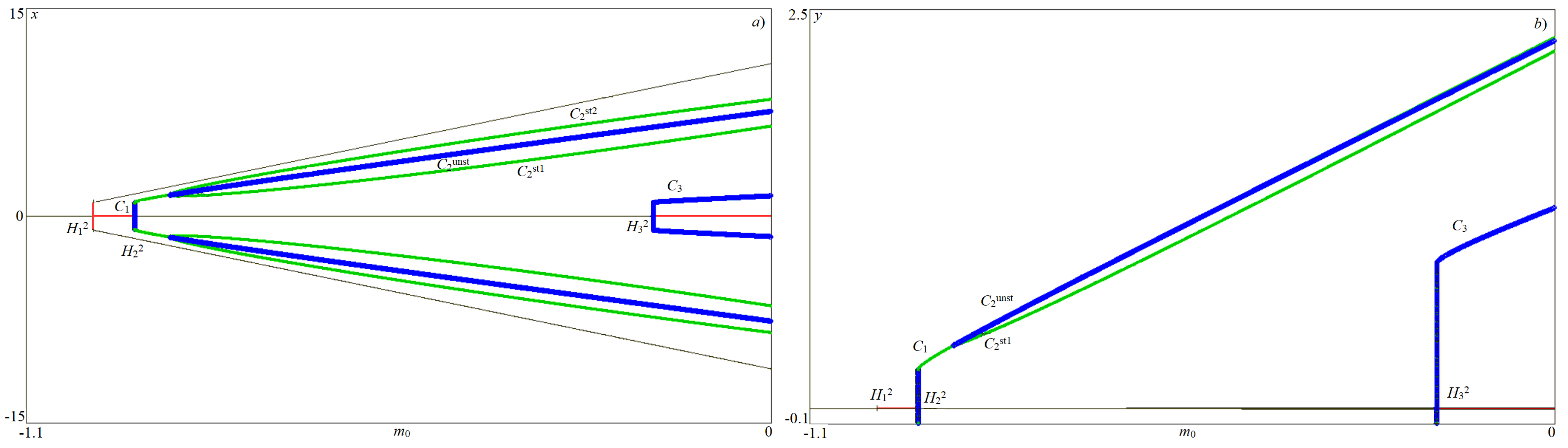}
\caption{Bifurcation diagram for the Chua system (\ref{Chua}) with $\alpha=8.4$, $\beta=12$, $\gamma=-0.005$, $m_1=-1.11$.}
\label{Fig.11}
\end{figure*}

Formation of hidden attractors in the area (\textbf{I}) in the parameter plane ($m_0$, $m_1$)
is as follows. In Fig.~\ref{Fig.11} the corresponding bifurcation diagrams are shown.
In this case it is better to exam the bifurcation diagrams by increasing
the parameter $m_0$. At $m_0 \approx -1.0004$ the Hopf bifurcation ($H_1^2$) occurs,
where the unstable zero equilibrium point becomes stable, simultaneously with pitchfork bifurcation,
and as a result of which two unstable equilibria are born. At $m_0 \approx -0.939$ the zero equilibrium point undergoes a supercritical Hopf bifurcation ($H_2^2$), and as a result the zero equilibrium point loses stability and a stable limit cycle $C_1$ is born, where it is situated between two symmetric unstable equilibria in projections onto the $x$ and $z$ variables,
where it surrounds all equilibria in a projection of $y$-variable.
Upon increasing the parameter $m_0$,
 the limit cycle undergoes a symmetry breaking bifurcation
 (it is the same as a pitchfork bifurcation),
 and splits into two stable symmetric limit cycles $C_2^{st1}$, $C_2^{st2}$
 and one unstable limit cycle $C_2^{unst}$.
 At $m_0 \approx -0.1761$ ($H_3^2$) the zero equilibrium point changes stability again.
 In this case the bifurcation is subcritical, and as a result the unstable limit cycle $C_3$ is born.
 The limit cycle $C_3$ forms boundaries of the basin of attraction of
 the zero stable equilibrium point.
 In Fig.~\ref{Fig.11}(b) is shown the projection of the $y$-variable.
 Thus, for $m_0>-0.1761$ ($H_3^2$) the limit cycles $C_2^{st1}$ and $C_2^{st2}$
 are isolated from all equilibria by the unstable limit cycle $C_3$,
 and these symmetric limit cycles $C_2^{st1}$ and $C_2^{st2}$ and chaotic attractors,
 which occur on the base of these cycles for another set of parameters, are hidden.

\section{Conclusion}

The dynamics of the Chua circuit gives a complex picture in the space
of controlling parameters.
The areas with similar behavior exist,
and the detailed study of the dynamics of the Chua system in these areas
allows one to reveal new hidden attractors.
It is shown that the formation of hidden attractors is connected with
the subcritical Hopf bifurcations of equilibrium points
and the saddle-node bifurcations of the limit cycles.
In general, the conjecture is that  for a globally bounded
autonomous system of ODE with asymptotically stable equilibrium point,
the subcritical Hopf bifurcation leads to the birth of a hidden attractor.
In two different areas of parameter plane it was found two types
of hidden attractors, namely, merged and separated attractors.
These features of hidden attractors are connected with the location
of stable and unstable equilibria and with the associated
with them unstable limit cycles in the phase space.
The open questions are what is the maximum number of coexisting attractors\footnote{
This question is related to the ``chaotic'' generalization \cite{LeonovK-2015-AMC}
of the second part of Hilbert's 16th problem
\emph{on the number and mutual disposition
of attractors and repellers in the chaotic multidimensional dynamical systems
and, in particular, their dependence on the degree of polynomials in the model};
see corresponding discussion, e.g. in \cite{SprottJKK-2017,ZhangChen-2017}.
}
that can be exhibited in the Chua system \eqref{Chua}
and how many of the coexisting attractors
can be hidden.

\nonumsection{Acknowledgments} \noindent This work was supported
by the Russian Science Foundation project 14-21-00041 (sections 2.2-4).
L.Chua's research is supported by grant no. AFOSR FA 9550-13-1-0136.


\end{document}